\documentstyle[12pt]{article}

\title{A determination of the
low energy parameters of the 2-d Heisenberg antiferromagnet}
\author{U.-J. Wiese$^{1,+}$ and H.-P. Ying$^{1,2}$ \\[2em]
$^1$ Institut f\"ur Theoretische Physik, Universit\"at Bern,\\
Sidlerstrasse 5, CH-3012 Bern, Switzerland \\[2em]
$^2$ Zhejiang Institute of Modern Physics, Zhejiang University, \\
Hangzhou 310027, P. R. China}

\begin{document}
\maketitle
\begin{abstract}

We perform numerical simulations of the 2-d antiferromagnetic quantum
Heisenberg model using an efficient cluster algorithm. Comparing the
finite size and finite temperature effects of various quantities with recent
results from chiral perturbation theory we are able to determine the low energy
parameters of the system very precisely. We find $e_0 = - 0.6693(1) J/a^2$
for the
ground state energy density, ${\cal M}_s = 0.3074(4)/a^2$ for the staggered
magnetization, $\hbar c = 1.68(1) J a$ for the spin wave velocity and
$\rho_s = 0.186(4) J$ for the spin stiffness. Our results agree with
experimental data for the undoped precursor insulators of high-$T_c$
superconductors.

\end{abstract}

\vspace{3cm}
$+_{\mbox{supported by the Schweizer Nationalfond}}$ \\
\newpage

The first high-$T_c$ superconductor to be discovered was
$\mbox{La}_{2-x}\mbox{Ba}_x\mbox{CuO}_4$ with $x \approx 0.15$ \cite{Bed86},
which has a layered structure with 2-d  copper-oxygen planes. The copper ions
are located
at the sites of a quadratic lattice with lattice spacing $a = 3.79$ \AA.
The undoped
material $\mbox{La}_2\mbox{CuO}_4$ is an insulator, however, with strong
antiferromagnetic interactions within the copper-oxygen planes
between electron spins localized at the
copper ions. The couplings between different layers are extremely weak.
Experimentally one observes long range antiferromagnetic order, i.e. a
spontaneous staggered magnetization ${\cal M}_s$ arises, which breaks the
$O(3)$ spin rotational symmetry down to $O(2)$. The low energy excitations
of the system are spinwaves (the so-called magnons) which are the Goldstone
bosons of the spontaneously broken $O(3)$ symmetry.
The physical situation can be modeled by the 2-d Heisenberg quantum spin system
with Hamiltonian
\begin{equation}
H = J \sum_{x,\mu} \vec{S}_x  \cdot \vec{S}_{x+\hat{\mu}},
\end{equation}
where $\vec{S}_x = \frac{1}{2}\vec{\sigma}_x$ is a spin $\frac{1}{2}$ operator
($\vec{\sigma}_x$ are Pauli matrices) located at the
point $x$ of a 2-d quadratic lattice with lattice spacing $a$.
The interaction is between nearest neighbors
($\hat{\mu}$ is the unit vector in $\mu$-direction) and $J > 0$ is the
antiferromagnetic exchange coupling. The question arises how well this model
describes the physics of the copper-oxygen planes in $\mbox{La}_2\mbox{CuO}_4$,
in particular how it compares quantitatively with experimental results.

Here we concentrate on the calculation of the low energy parameters of the
model, which determine the dynamics of the Goldstone bosons. These are the
staggered magnetization ${\cal M}_s$, the spinwave velocity $\hbar c$ and
the spin stiffness $\rho_s$. Based on symmetry considerations chiral
perturbation theory makes very strong predictions for the magnon dynamics,
containing the low energy parameters as the only unknown constants.
Recently, Hasenfratz and Niedermayer have worked out the chiral perturbation
theory for the antiferromagnet in great detail up to two-loop order
\cite{Has92}. Lower order results had been obtained before by Fisher
\cite{Fis89} and by Neuberger and Ziman \cite{Neu89}.
Here we only quote the results of ref.\cite{Has92}
that are essential for our study. We consider
the system at finite temperature $T$ and in a finite spatial volume
of size $L \times L$ with periodic boundary conditions, with $l^3 =
\hbar c / T L$ such that $l$ is of order 1. For small enough temperatures
$T \ll 2 \pi \rho_s$ and large enough volumes $\hbar c / L \ll 2 \pi \rho_s$
Hasenfratz and Niedermayer obtained the following results: for the internal
energy density
\begin{equation}
e(T,L) = e_0 - \frac{T}{3 L^2} \left\{1 + l \frac{d}{dl} \beta_0(l) -
\frac{\hbar c}{\rho_s L l}\left[\beta_1(l) - l \frac{d}{dl}
\beta_1(l)\right] + ...\right\},
\end{equation}
where $e_0$ is the ground state energy density; for the staggered
susceptibility
\begin{equation}
\chi_s(T,L) = \frac{{\cal M}_s^2 L^2}{3 T} \left\{1 +
2 \frac{\hbar c}{\rho_s L l}
\beta_1(l) + \left(\frac{\hbar c}{\rho_s L l}\right)^2 \left[\beta_1(l)^2 +
3 \beta_2(l)\right] + ...\right\};
\end{equation}
and for the uniform susceptibility
\begin{equation}
\chi(T,L) = \frac{2 \rho_s}{3 (\hbar c)^2} \left\{1 + \frac{1}{3}
\frac{\hbar c}{\rho_s L l} \tilde{\beta}_1(l) +
\frac{1}{3} \left(\frac{\hbar c}{\rho_s L l}\right)^2
\left[\tilde{\beta}_2(l) - \frac{1}{3} \tilde{\beta}_1(l)^2 - 6 \psi(l)
\right] + ...\right\}.
\end{equation}
The functions $\beta_i(l)$, $\tilde{\beta}_i(l)$ and $\psi(l)$ are shape
coefficients which depend only on $l$ and which are described in detail in
ref.\cite{Has92}. We will use these results
of chiral perturbation theory to determine the unknown low energy parameters
$e_0$, ${\cal M}_s$, $\hbar c$ and $\rho_s$ from a fit of $e(T,L)$,
$\chi_s(T,L)$ and $\chi(T,L)$ to numerical Monte Carlo data. This method has
been used before for classical spin models and for relativistic quantum field
theories \cite{Has89}.

First we decompose the Hamiltonian into
$H = H_1 + H_2 + H_3 + H_4$ with
\begin{eqnarray}
&& H_1 = J \sum_{x = (2m,n)} \vec{S}_x \cdot \vec{S}_{x+\hat{1}}, \,\,\,
H_2 = J \sum_{x = (m,2n)} \vec{S}_x \cdot \vec{S}_{x+\hat{2}}, \nonumber \\
&& H_3 = J \sum_{x = (2m+1,n)} \vec{S}_x  \cdot \vec{S}_{x+\hat{1}}, \,\,\,
H_4 = J \sum_{x = (m,2n+1)} \vec{S}_x  \cdot \vec{S}_{x+\hat{2}},
\end{eqnarray}
and we use the Suzuki-Trotter formula for the partition function
\begin{equation}
Z = \mbox{Tr} \exp(- \beta H) = \lim_{N \rightarrow \infty}
\mbox{Tr}[\exp(- \epsilon \beta H_1) \exp(- \epsilon \beta H_2)
\exp(- \epsilon \beta H_3) \exp(- \epsilon \beta H_4)]^N,
\end{equation}
where $\beta = 1/T$ is the inverse temperature and
$\epsilon = 1/N$ determines the lattice spacing in the euclidean time
direction. By inserting complete sets of eigenstates
$|1\rangle$ and $|\mbox{--1}\rangle$ of $\sigma_x^3$
between the factors $\exp(- \epsilon \beta H_i)$ we map the 2-d quantum spin
system to a 3-d induced classical system of Ising-like variables $s(x,t) =
\pm 1$ ($t$ labels the euclidean time slice)
\begin{equation}
Z = \prod_{x,t} \sum_{s(x,t) = \pm 1} \exp(- S)
\end{equation}
with an action
\begin{eqnarray}
S & = &
\sum_{x=(2m,n),t=4p} S[s(x,t),s(x+\hat{1},t),s(x,t+1),s(x+\hat{1},t+1)]
\nonumber \\ & + &
\sum_{x=(m,2n),t=4p+1}
S[s(x,t),s(x+\hat{2},t),s(x,t+1),s(x+\hat{2},t+1)]
\nonumber \\ & + &
\sum_{x=(2m+1,n),t=4p+2}
S[s(x,t),s(x+\hat{1},t),s(x,t+1),s(x+\hat{1},t+1)]
\nonumber \\ & + &
\sum_{x=(m,2n+1),t=4p+3}
S[s(x,t),s(x+\hat{2},t),s(x,t+1),s(x+\hat{2},t+1)].
\end{eqnarray}
The classical spins interact with each other via four-spin couplings
$S[s(x,t),s(x+\hat{\mu},t),s(x,t+1),s(x+\hat{\mu},t+1)]$ associated with
time-like
plaquettes. Up to a trivial additive constant one has
$S[1,1,1,1] =$ $S[-1,-1,-1,-1] = 0$, $S[1,-1,1,-1] =$ $S[-1,1,-1,1] =$
$- \log[\frac{1}{2}(\exp(\epsilon \beta J) + 1)]$ and
$S[1,-1,-1,1] = $ $S[-1,1,1,-1] =$
$- \log[\frac{1}{2}(\exp(\epsilon \beta J) - 1)]$.
All other action values are
infinite. This causes problems in numerical simulations because many spin
configurations are forbidden and the updating must respect several constraints.
In a previous paper we have introduced blockspins \cite{Wie92}
to resolve the constraints. For the 1-d antiferromagnetic spin chain the
blockspin model is not frustrated and the use of a blockspin cluster algorithm
eliminates critical slowing down. In two dimensions, however, frustration
causes severe problems. Recently, Evertz, Lana and Marcu \cite{Eve92} have
developed loop cluster algorithms for vertex models, which can also be applied
to quantum spin systems. The algorithm constructs closed loops of spins and
flips them simultaneously.
The loop cluster algorithm does not suffer from
frustration but it may suffer from so-called freezing. Freezing occurs when a
loop branches out many times and fills a large fraction of the whole volume. We
find that freezing does not arise for the
Heisenberg antiferromagnet. This is essential for the success of our numerical
study.

The algorithm constructs
loops by first selecting a starting point $(x,t)$ at random.
The spin $s(x,t)$ participates in two plaquette interactions,
one at euclidean times before and one at euclidean times after $t$. When
$s(x,t) = 1$ we consider the plaquette
interaction at the later time, and for $s(x,t) = -1$
we consider the interaction at the earlier time. The
corresponding plaquette configuration
is characterized by the spin orientations at the four corners.
One of the corners will be the next point on the loop.
For configurations $C_1 = [1,1,1,1]$ or $[-1,-1,-1,-1]$
the next point is the time-like nearest neighbor of $(x,t)$ on the plaquette.
For configurations $C_2 = [1,-1,1,-1]$ or $[-1,1,-1,1]$
the next point on the loop is
with probability $p = 2/(\exp(\epsilon \beta J) + 1)$ the time-like nearest
neighbor, and with probability $1 - p$ the
space-like nearest neighbor of $(x,t)$. Finally, for configurations
$C_3 = [1,-1,-1,1]$ or $[-1,1,1,-1]$ the next point on the loop is
the space-like nearest neighbor of $(x,t)$.  Once the next point on the loop
is determined the process is repeated until the loop closes. Then all spins on
the loop are flipped simultaneously. The algorithm obeys detailed balance,
$p(C_i) w(C_i \rightarrow C_j) = p(C_j) w(C_j \rightarrow C_i)$,
where $p(C_1) = 1$, $p(C_2) = \frac{1}{2}(\exp(\epsilon \beta J) + 1)$,
$p(C_3) = \frac{1}{2}(\exp(\epsilon \beta J) - 1)$ and
$w(C_i \rightarrow C_j)$ is the
transition probability to go from a plaquette configuration $C_i$ to $C_j$.
Indeed one has
\begin{eqnarray}
&&p(C_1) w(C_1 \rightarrow C_2) = 1 = \frac{1}{p} \, p =
p(C_2) w(C_2 \rightarrow C_1), \nonumber \\
&&p(C_2) w(C_2 \rightarrow C_3) = \frac{1}{p} (1 - p) =
\frac{1}{2}(\exp(\epsilon \beta J) - 1) = p(C_3) w(C_3 \rightarrow C_2).
\end{eqnarray}
In our construction a loop cannot branch out and hence freezing does not arise.
Cluster algorithms offer the possibility to use improved estimators which
reduce
the variance of different observables. For example, the
uniform susceptibility can be
expressed as
$\chi a^2 J = \frac{\beta J}{4 N} \langle M_{\cal C}^2/|{\cal C}|\rangle$,
where $4N$ is the number of points in the euclidean time direction,
$|{\cal C}| = \sum_{(x,t) \in {\cal C}} 1$ is the size of the loop ${\cal C}$
and $M_{\cal C} = \frac{1}{2} \sum_{(x,t) \in {\cal C}} s(x,t)$
is the loop magnetization. It is interesting to note that
clusters with nonzero magnetization
must wrap around the lattice in the euclidean time direction. Small clusters
which do not wrap around the lattice
have $M_{\cal C} = 0$. Similarly, one can
define improved estimators for the staggered susceptibility
$\chi_s$ and for the internal energy density $e$.

Some results of our numerical simulations are collected in table 1. We have
performed measurements for
three inverse temperatures $\beta J = 5,10,15$ and for
different spatial sizes $L/a = 6,8,...,20$. We have always performed 10000
loop updates for equilibration followed by 100000 measurements using the
improved estimators. The autocorrelation times of the loop cluster algorithm
are at most a few sweeps, and we see no indication of critical slowing down.
With standard local algorithms it would be impossible to reach temperatures
as low as the ones we use here, because of severe problems with
slowing down.
\begin{table} \begin{center}
\caption{Numerical data for $e$, $\chi_s$ and $\chi$. $\newline \;$}
\begin{tabular}{|c|c|c|c|c|c|} \hline
$\beta J$ & $L/a$ & $4N$ & $e a^2/J$ & $\chi_s a^2 J$ & $\chi a^2 J$ \\ \hline
 5 &  6 & 256 & -0.678(1) &  9.67(3) & 0.0482(3) \\ \hline
 5 &  8 & 256 & -0.673(1) & 16.08(5) & 0.0514(3) \\ \hline
 5 & 10 & 256 & -0.672(1) & 23.73(7) & 0.0527(3) \\ \hline
 5 & 12 & 256 & -0.671(1) &  32.3(1) & 0.0530(3) \\ \hline
 5 & 14 & 256 & -0.671(1) &  41.7(1) & 0.0519(3) \\ \hline
 5 & 16 & 256 & -0.669(1) &  52.6(2) & 0.0531(3) \\ \hline
 5 & 18 & 256 & -0.672(1) &  64.3(2) & 0.0528(3) \\ \hline
 5 & 20 & 256 & -0.670(1) &  76.3(3) & 0.0535(3) \\ \hline
10 &  6 & 512 & -0.679(1) & 14.64(5) & 0.0268(3) \\ \hline
10 &  8 & 512 & -0.675(1) &  27.5(1) & 0.0406(3) \\ \hline
10 & 10 & 512 & -0.673(1) &  42.9(2) & 0.0442(3) \\ \hline
10 & 12 & 512 & -0.673(1) &  60.6(2) & 0.0460(3) \\ \hline
10 & 14 & 512 & -0.670(1) &  81.2(3) & 0.0469(3) \\ \hline
10 & 16 & 512 & -0.673(1) & 103.1(4) & 0.0476(3) \\ \hline
10 & 18 & 512 & -0.672(1) & 129.0(4) & 0.0480(3) \\ \hline
10 & 20 & 512 & -0.671(1) & 156.2(5) & 0.0477(3) \\ \hline
15 &  6 & 768 & -0.681(1) & 16.71(6) & 0.0111(3) \\ \hline
15 &  8 & 768 & -0.675(1) &  34.8(1) & 0.0287(3) \\ \hline
15 & 10 & 768 & -0.674(1) &  57.9(2) & 0.0385(3) \\ \hline
15 & 12 & 768 & -0.674(1) &  83.8(3) & 0.0420(3) \\ \hline
15 & 14 & 768 & -0.672(1) & 113.6(4) & 0.0439(3) \\ \hline
15 & 16 & 768 & -0.671(1) & 148.0(5) & 0.0451(3) \\ \hline
15 & 18 & 768 & -0.670(1) & 187.0(6) & 0.0457(3) \\ \hline
15 & 20 & 768 & -0.671(1) & 227.6(8) & 0.0457(3) \\ \hline
\end{tabular} \end{center} \end{table}
In table 1 the lattice spacing has been fixed to $\epsilon \beta J =
\frac{5}{64}$.
We have also performed runs on coarser lattices with
$\epsilon \beta J = \frac{5}{32}$ and $\frac{5}{48}$.
This allows us to extrapolate our data to
the euclidean time continuum limit $\epsilon \rightarrow 0$. After the
extrapolation we fit the results to the above expressions
from chiral perturbation theory. The data for $e$, $\chi_s$ and $\chi$ are all
fitted simultaneously. Our best fit with $\chi^2/\mbox{dof} = 1.4$
is shown in fig 1. The finite size and finite temperature effects of the
internal energy density depicted in fig.1a are very small (of the order of our
statistical errors), while the effects on the susceptibilities are much larger.
For low temperature and small volume some data have been excluded from the fit
because for them $l$ is not of order 1. The fit gives the following values for
the low energy parameters
\begin{equation}
e_0 = -0.6693(1) J/a^2, \,\, {\cal M}_s = 0.3074(4)/a^2, \,\,
\hbar c = 1.68(1) J a,
\,\, \rho_s = 0.186(4) J.
\end{equation}
To our knowledge this is the most accurate determination of
these zero temperature and
infinite volume properties from a simulation of the partition function
at finite temperature and finite volume.
The result for the ground state energy density agrees with
different zero temperature Monte Carlo calculations \cite{Bar91} which yield
$e_0 = -0.6692(1) J/a^2$. Our results are consistent
with an analytic expansions around the Ising limit \cite{Zhe91} which gives
$e = 0.6693(1) J/a^2$ and ${\cal M}_s = 0.307(1)/a^2$,
but not consistent with a recent
large scale numerical study using a standard local algorithm
\cite{Mak91} which obtained $\rho_s = 0.199(2) J$.
Finally, we compare our results with experimental data.
Using inelastic neutron
scattering the spin wave velocity $\hbar c = 0.85(3) \mbox{eV}$\AA \, has been
measured \cite{Aep89}, while an analysis of Raman scattering data \cite{Sin89}
yields $J = 0.128(6) \mbox{eV} = 1480(70) \mbox{K}$.
Using this together with $a = 3.79$\AA \, the
experiments on $\mbox{La}_2\mbox{CuO}_4$ obtain $\hbar c = 1.75(9) J a$.
This is
consistent with our result for the Heisenberg antiferromagnet. Using the
experimental values for the spinwave velocity and for the lattice spacing
we obtain an independent estimate of the exchange coupling in
$\mbox{La}_2\mbox{CuO}_4$
\begin{equation}
J = 0.133(5) \mbox{eV} = 1540(60) \mbox{K}.
\end{equation}
The agreement between our numerical results and the predictions
of chiral perturbation theory confirms that the Heisenberg model has long
range antiferromagnetic order, and that its low energy dynamics is dominated
by magnons. A precise determination of the low energy parameters that
determine the magnon physics was possible only because the loop cluster
algorithm is very efficient also at low temperatures.
Recently, a loop cluster algorithm has
been constructed for lattice fermion systems \cite{Wie92b}.
This raises hopes that
numerical investigations  of similar accuracy
become feasible for the Hubbard model, and hence for
high-$T_c$ superconductors like $\mbox{La}_{2-x}\mbox{Ba}_x\mbox{CuO}_4$.

We are indebted to P. Hasenfratz and F. Niedermayer for
making their results available to us prior to publication. We also like to
thank them for many interesting discussions about quantum antiferromagnets.

\section*{Figure Caption}

Fig.1: The fit of the Monte-Carlo data for the internal energy $e a^2/J$ (a),
the staggered susceptibility $\chi_s a^4 J/L^2$ (b), and the uniform
susceptibility $\chi a^2 J$ (c). The dots, squares and triangles are the
Monte-Carlo data for $\beta J = 5$, 10 and 15 respectively. The corresponding
fit functions are represented by the solid, dashed and dotted curves.


\newpage

\begin{thebibliography}{13}
\bibitem{Bed86}
J. G. Bednorz and K. A. M\"uller, Z. Phys. B64 (1986) 189.
\bibitem{Has92}
P. Hasenfratz and F. Niedermayer, Bern preprint 1992, BUTP-92/46.
\bibitem{Fis89}
D. S. Fisher, Phys. Rev. B39 (1989) 11783.
\bibitem{Neu89}
H. Neuberger and T. Zimann, Phys. Rev. B39 (1989) 2608.
\bibitem{Has89}
A. Hasenfratz, K. Jansen, J. Jers\'{a}k, C. B. Lang, H. Leutwyler and
T. Neuhaus,
Z. Phys. C46 (1989) 257; \\
I. Dimitrovi\'{c}, P. Hasenfratz, J. Nager and F. Niedermayer, Nucl. Phys. B350
(1991) 893.
\bibitem{Wie92}
U.-J. Wiese and H.-P. Ying, Phys. Lett. A168 (1992) 143.
\bibitem{Eve92}
H. G. Evertz and M. Marcu,
talk presented at the 1992 Symposium on Lattice Field Theory,
Amsterdam, 15-19 September 1992, Tallahassee preprint (1992),
FSU-SCRI-92-165; \\
H. G. Evertz, G. Lana and M. Marcu, Tallahassee preprint (1992),
FSU-SCRI-92-164.
\bibitem{Bar91}
T. Barnes, Int. J. Mod. Phys. C2 (1991) 659.
\bibitem{Zhe91}
W. Zheng, J. Oitmaa and C. J. Hamer, Phys. Rev. B43 (1991) 8321.
\bibitem{Mak91}
M. S. Makivi\'{c} and H.-Q. Ding, Phys. Rev. B43 (1991) 3562.
\bibitem{Aep89}
G. Aeppli, S. M. Hayden, H. A. Mook, Z. Fisk, S.-W. Cheong, D. Rytz,
J. Remeika, G. P. Espinosa and A. S. Cooper, Phys. Rev. Lett. 62 (1989) 2052.
\bibitem{Sin89}
R. R. P. Singh et. al., Phys. Rev. Lett. 62 (1989) 2736.
\bibitem{Wie92b}
U.-J. Wiese, Bern preprint 1992, BUTP-92/45.
\end{thebibliography}
\end{document}